\documentclass[aoas,MSNbibl,nameyear,seceqn,dvips]{arximspdf}
\usepackage{graphicx}
\doi{10.1214/13-AOAS690} 
\volume{8}
\issue{1}
\pubyear{2014}
\firstpage{352}
\lastpage{376}

\makeatletter

\newcommand{\rright}{\right}
\newcommand{\lleft}{\left}
\newcommand{\bs}[1]{\bolds{#1}}
\newcommand{\mb}[1]{\mathbf{#1}}

\def\bX{\mathbf{X}}
\def\bS{\mathbf{S}}
\def\bY{\mathbf{Y}}

\def\bU{\mathbf{U}}
\def\bC{\mathbf{C}}
\def\bbeta{\bolds{\beta}}
\def\balpha{\bolds{\alpha}}

\def\bmuhat{\hat{\bolds{\mu}}}
\makeatother

\begin{document}
\begin{frontmatter}

\title{Joint analysis of SNP and gene expression data in genetic
association studies of complex diseases\thanksref{T1}}
\runtitle{Joint analysis of SNP and expression in association studies}

\begin{aug}
\author[A]{\fnms{Yen-Tsung} \snm{Huang}\corref{}\thanksref{m1}\ead[label=e1]{Yen-Tsung\_Huang@brown.edu}},
\author[B]{\fnms{Tyler J.} \snm{VanderWeele}\thanksref{m2}\ead[label=e2]{tvanderw@hsph.harvard.edu}}
\and
\author[C]{\fnms{Xihong} \snm{Lin}\thanksref{m2}\ead[label=e3]{xlin@hsph.harvard.edu}}
\runauthor{Y.-T. Huang, T.~J. VanderWeele and X. Lin}
\affiliation{Brown University\thanksmark{m1} and Harvard
University\thanksmark{m2}}
\address[A]{Y.-T. Huang\\
Department of Epidemiology\\
Brown University\\
121 South Main Street\\
Providence, Rhode Island 02912\\
USA\\
\printead{e1}} 
\address[B]{T. J. VanderWeele\\
Departments of Epidemiology and Biostatistics\\
Harvard School of Public Health\\
665 Huntington Avenue\\
Boston, Massachusetts 02115\\
USA\\
\printead{e2}}
\address[C]{X. Lin\\
Departments of Biostatistics\\
Harvard School of Public Health\\
665 Huntington Avenue\\
Boston, Massachusetts 02115\\
USA\\
\printead{e3}}
\end{aug}
\thankstext{T1}{Supported in part by grants from the National Cancer
Institute (R37-CA076404 and \mbox{P01-CA134294}) and the National
Institute of Environmental Health P42-ES016454.}

\received{\smonth{1} \syear{2013}}
\revised{\smonth{8} \syear{2013}}

%
\begin{abstract}
Genetic association studies have been a popular approach for assessing
the association between common Single Nucleotide Polymorphisms (SNPs)
and complex diseases. However, other genomic data involved in the
mechanism from SNPs to disease, for example, gene expressions, are
usually neglected in these association studies. In this paper, we
propose to exploit gene expression information to more powerfully test
the association between SNPs and diseases by jointly modeling the
relations among SNPs, gene expressions and diseases. We propose a
variance component test for the total effect of SNPs and a gene
expression on disease risk. We cast the test within the causal
mediation analysis framework with the gene expression as a potential
mediator. For eQTL SNPs, the use of gene expression information can
enhance power to test for the total effect of a SNP-set, which is the
combined direct and indirect effects of the SNPs mediated through the
gene expression, on disease risk. We show that the test statistic under
the null hypothesis follows a mixture of $\chi^2$ distributions, which
can be evaluated analytically or empirically using the resampling-based
perturbation method. We construct tests for each of three disease
models that are determined by SNPs only, SNPs and gene expression, or
include also their interactions. As the true disease model is unknown
in practice, we further propose an omnibus test to accommodate
different underlying disease models. We evaluate the finite sample
performance of the proposed methods using simulation studies, and show
that our proposed test performs well and the omnibus test can almost
reach the optimal power where the disease model is known and correctly
specified. We apply our method to reanalyze the overall effect of the
SNP-set and expression of the \textit{ORMDL3} gene on the risk of asthma.
\end{abstract}

%

\end{frontmatter}

\section{Introduction}\label{sec1}
\label{sintro}

Genome-wide association studies (GWAS) constitute a popular approach
for investigating the association of common Single Nucleotide
Polymorphisms (SNPs) with complex diseases. Usually, a large number of
SNP \mbox{markers} are tested across the genome. Great interest lies in
improving power of testing SNP effects by borrowing additional
biological information. Indeed, a major criticism of genetic
association studies lies in its agnostic style [\citet{hunter}]:
none of the biological knowledge was encoded in the standard genetic
association analyses. To overcome such limitations, multi-marker
analysis has been advocated to integrate biological information into
statistical analyses and to decrease the number of tests [\citet
{kwee}; \citet{wu}]. Analysis using SNP-sets grouped
by physical locations has better performance than the standard single
SNP analysis in reanalyzing the breast cancer GWAS data [\citet
{wu}]. SNPs can also be grouped into a SNP-set according to biological
pathways, in which a gene harmonizes with other genes to exert
biological functions.

The two factors we try to bridge in genetic association studies are
SNPs and disease risk. Despite the success of SNP-set analyses in
assembling multiple SNPs based on biological information, mechanistic
pathways between SNPs (SNP-sets) and disease are still neglected. Given
the availability of multiple sources in genomic data (e.g., gene
expression and SNPs) [\citet{moffatt}; \citet{cusanovich}],
it is desirable to perform joint analysis by integrating multiple
sources of genomic data. Here we combine the information of SNPs and
gene expression by introducing gene expression as a mediator in the
causal pathway from SNPs to disease. \textit{Biologically}, gene
expression can be determined by the DNA genotype [\citet{morley};
\citet{cheung}; \citet{fu}] and that gene expression can also
affect disease risk [\citet{dermitzakis}]. Moreover, results from
the SNP-set analysis augmented by a biological model can be more
scientifically meaningful. \textit{Statistically}, gene expression can
help explain variability of the effect of SNPs on disease when there
exists an effect of SNPs on disease via gene expression and thus
increases the power of detecting the overall effect of SNPs on disease risk.

SNPs that regulate mRNA expression of a gene are so-called expression
Quantitative Trait Loci (eQTL) [\citet{schadt2003}].
Statistically, eQTL SNPs can be viewed as the SNPs that are correlated
with mRNA expression of a gene. \textit{Cis}-eQTL SNPs are the SNPs that
are within or around the corresponding gene, and \textit{trans}-eQTL SNPs
are those that are far away or even on different chromosomes. Numerous
genome-wide eQTL analyses have been reported to comprehensively capture
such a DNA--RNA (i.e., SNPs-gene expression) association in the genome
in different tissues and organisms [\citet{schadt2003}; \citet
{morley}; \citet{innocenti}]. eQTL results can be external
information to prioritize the discovery of susceptibility loci in
genome-wide association studies [\citet{hsu}; \citet{zhong};
\citet{zhang}]. Methods are available to integrate multiple
genomic data to draw causal inference on a biological network
[\citet{schadt2005}; \citet{zhu}; \citet{hageman};
\citet{neto}]. We focus in this paper on \textit{joint analysis} of
multiple eQTL SNPs of a gene and their corresponding mRNA expression
for their effects on disease phenotypes. Compared with multi-SNP
analyses, this approach further incorporates eQTLs into genetic
association studies and accounts for a biological process (from DNA to
RNA) within a gene to improve power.

\begin{figure}[b]
\includegraphics{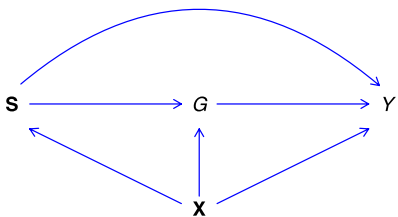}
\caption{Causal diagram of the mediation model. $\mb{S}$ is a set of correlated
exposures, for example, SNP set; $G$ is a mediator, for example, gene
expression; $Y$ is an outcome, for example, disease (yes/no); and $\mb{X}$ are
covariates, including the true and potential confounders.}\label{fig1}
\end{figure}

This paper is motivated by an asthma genome-wide association study of
subjects of British descent (MRC-A), in which the association between
SNPs at the \mbox{\textit{ORMDL3}} gene and the risk of childhood asthma was
investigated [\citet{dixon}; \citet{moffatt}]. The MRC-A data
set consists of 108 cases and 50 controls with both SNP genotype
(Illumina 300K) and gene expression (Affymetrix HU133A 2.0) data
available. The original genome-wide study reported that the 10 typed
SNPs on chromosome 17q21 where \textit{ORMDL3} is located were strongly
associated with childhood asthma in MRC-A data, and the results were
validated in several other independent studies. The authors also found
that each of these 10 SNPs was highly correlated with gene expression
of \textit{ORMDL3}, which is also associated with asthma. The 10 SNPs,
\textit{ORMDL3} expression and asthma status can be illustrated as the
$\mb{S}$, $G$ and $Y$, respectively, in Figure~\ref{fig1}. Instead of analyzing
SNP-expression, expression-asthma and SNP-asthma associations
separately and univariately, here we are interested in assessing the
overall genetic effect of \textit{ORMDL3} on the occurrence of
childhood asthma, by jointly analyzing SNP and gene expression data and
accounting for the possibility that the \textit{ORMDL3} gene expression
might be a causal mediator for the association of the SNPs in the
\textit{ORMDL3} gene and asthma risk. Our ultimate goal is to integrate
multiple sources of genomic data for genetic association analyses.

%

In this paper we propose to jointly model a set of SNPs within a gene,
a gene expression and disease status, where a logistic model is used to
model the dependence of disease status on the SNP-set and the gene
expression, and a linear model is used for the dependence of the gene
expression on the SNP-set, both adjusting for covariates. We are
primarily interested in testing whether a gene, whose effects are
captured by SNPs and/or gene expression, is associated with a disease
phenotype. We formulate this hypothesis in the causal mediation
analysis framework [\citet{robins1992}; \citeauthor{vanderweele2009}
(\citeyear{vanderweele2009,vanderweele2010}); \citet{imai}]. Note that the previous
work on causal mediation analysis is mostly focused on estimation.

We use the joint model to derive the direct and indirect effects of a
SNP-set mediated through gene expression on disease risk. For eQTL
SNPs, we show that the total effect of a gene on a disease captured by
a set of SNPs and a gene expression corresponds to the total effects of
the SNP-set, which are the combined direct effects and indirect effects
of the SNPs mediated through the gene expression, on a disease. This
framework allows us to study how the use of gene expression data can
enhance power to test for the total effect of a SNP-set on disease
risk. We study the impact of model misspecification using the
conventional SNP-only models when the true model is that both the SNPs
and the gene expression affect the disease outcome. For non-eQTL SNPs,
the null hypothesis simply corresponds to the joint effects of the SNPs
and the gene expression.

Due to a potentially large number of SNPs within a gene and some of
them possibly being highly correlated, that is, in high linkage
disequilibrium (LD), conventional tests, such as the likelihood ratio
test, for the total effects of multiple SNPs and a gene expression, do
not perform well. We propose in this paper a~variance component test to
assess the overall effects of a SNP set and a gene expression on
disease risk, under the null that the test statistic follows a mixture
of $\chi^2$ distributions, which can be approximated analytically or
empirically using a resampling-based perturbation procedure [\citet{parzen};
\citet{cai}]. As the true disease model is often
unknown, we construct an omnibus test to improve the power by
accommodating different underlying disease models.

The rest of the paper is organized as follows. In Section~\ref{sec2} we
introduce the joint model for SNPs, a gene expression and disease as
well as the null hypothesis of no joint effect of the SNPs and the gene
expression on a disease phenotype. In Section~\ref{sec3} we propose a variance
component score test for the total effect of SNPs and gene expression,
and construct an omnibus test to maximize the test power across
different underlying disease models. In Section~\ref{sec4} we interpret the null
hypothesis and study the assumptions within the framework of causal
mediation modeling for eQTL SNPs and non-eQTL SNPs. In Section~\ref{sec5} we
evaluate the finite sample performance of the proposed test using
simulation studies and show that the omnibus test is robust and
performs well in different situations. In Section~\ref{sec6} we apply the
proposed method to study the overall effect of the \textit{ORMDL3} gene
contributed by both the SNPs and the gene expression on the risk of
childhood asthma, followed by discussions in Section~\ref{sec7}.

\section{The model and the null hypothesis}\label{sec2}\label{smodel}

The statistical problem is to jointly model the effect of a set of SNPs
and a gene expression on the occurrence of a~disease. Assume for
subject $i$ ($i=1,\ldots,n$) an outcome of interest $Y_i$ is dichotomous
(e.g., case/control), whose mean is associated with $q$ covariates ($\mb
{X}_i$, with the first covariate to be 1, i.e., the intercept), $p$
SNPs in a SNP-set ($\mb{S}_i$), mRNA expression of a gene ($G_i$) and
possibly the interactions between the SNPs and the gene expression as
\begin{equation}
\operatorname{logit}\bigl\{P(Y_i=1|\mb{S}_i,
G_i, \mb{X}_i)\bigr\} =\mb{X}_i^T
\bs{\alpha}+\mb{S}_i^T\bs{\beta}_S+G_i
\beta_G+ G_i\mb{S}_i^T\bs{
\gamma}, \label{ymodel}
\end{equation}
where $\bs{\alpha}=(\alpha_1,\ldots, \alpha_q)^T, \bs{\beta}_S=(\beta
_{S_1},\ldots, \beta_{S_p})^T, \beta_G$, and $\bs{\gamma}=(\gamma_1,\ldots,
\gamma_p)^T$ are the regression coefficients for the covariates, the
SNPs, the gene expression, and the interactions of the SNPs and the
gene expression, respectively. A SNP-set and gene expression pair can
be defined in multiple ways. For example, $\bS$ can be the SNPs in a
gene and $G$ is the mRNA expression of the gene. In the asthma data
example, $\bS$ are the 10 typed SNPs around \textit{ORMDL3} and $G$ is
the expression of \textit{ORMDL3}. Alternatively, one can choose the
SNP-set/expression pair based on the eQTL study: eQTL
SNP-set/corresponding gene expression.

As SNPs can affect gene expression [\citet{schadt2003}; \citet
{morley}; \citet{innocenti}], for each subject $i$, we next
consider a linear model for the continuous gene expression $G_i$ (i.e.,
the mediator), which depends on the $q$ covariates ($\mb{X}_i$) and the
$p$ SNPs ($\mb{S}_i$):
\begin{equation}
G_i=\mb{X}_i^T\bs{\phi}+
\mb{S}_i^T\bs{\delta}+\varepsilon_{i},
\label{gmodel}
\end{equation}
where $\bs{\phi}=(\phi_1,\ldots, \phi_q)^T$ and $\bs{\delta}=(\delta_1,\ldots, \delta_p)^T$ are the regression coefficients for the covariates
and the SNPs, respectively, and $\varepsilon_{i}$ follows a normal
distribution with mean $0$ and variance $\sigma_G^2$. Again, the $p$
SNPs can be the SNPs within a gene or the eQTL SNPs corresponding to a
gene for which the expression level is measured.

Our goal is to test for the total effect of a gene captured by the SNPs
in a set $\bS$ and a gene expression $G$ on $Y$, which can be written
using the regression coefficients in model (\ref{ymodel}) as
\begin{equation}
H_0\dvtx  \bs{\beta}_S=\mb{0},\qquad \beta_G=0,\qquad \bs{\gamma}=\mb{0}. \label{null}
\end{equation}
Note that the null hypothesis (\ref{null}) only involves the parameters
in the $[Y|\bS,G, \bX]$ model (\ref{ymodel}). We use the $[G|\bS,\bX]$
model in Section~\ref{sinterpret} to facilitate interpretation of the
null hypothesis (\ref{null}) and to study the assumptions within the
causal mediation analysis framework. Throughout the paper, we term this
null hypothesis as a test for the \textit{total effect of a gene}. Later,
in Section~\ref{sinterpret} we will show that it corresponds to the
\textit{total effect of SNPs} for eQTL SNPs and simply to the joint effect
of SNPs and expression for non-eQTL SNPs.

\section{Test for the total effects of a gene}\label{sec3}
\label{stest}
\subsection{Test statistic for the total effect of a gene}\label{sec3.1}
\label{ssvct}

We propose in this section to test for the null hypothesis of no total
effect of a gene (\ref{null}) under model (\ref{ymodel}). As the number
of SNPs ($p$) in a gene might be large and some might be highly
correlated (due to linkage disequilibrium), the likelihood ratio test
(LRT) or multivariate Wald test for the null hypothesis (\ref{null})
uses a large degree of freedom (DF) and has limited power. To overcome
this problem, we assume under model (\ref{ymodel}) the regression
coefficients of the individual main SNP effects ${\beta}_{Sj}$ are
independent and follow an arbitrary distribution with mean 0 and
variance $\tau_S$, and the SNP-by-expression interaction coefficients
${\gamma}_j$ ($j=1,\ldots,p$) are independent and follow another arbitrary
distribution with mean 0 and variance $\tau_I$. The resulting outcome
model (\ref{ymodel}) hence becomes a logistic mixed model. The problem
for testing the null hypothesis (\ref{null}) becomes a joint test of
variance components ($\tau_S=\tau_I=0$) and the scalar regression
coefficient for the fixed gene expression effect ($\beta_G=0$) in the
induced logistic mixed models as $H_0\dvtx \tau_S=\tau_I=0$ and $\beta_G=0$.
One can easily show that the scores for $\tau_S$, $\beta_G$ and
$\tau_I$ under the induced logistic mixed models are
\begin{eqnarray*}
U_{\tau_S}&=&\{\mb{Y}-\bmuhat_0\}^T
\mathbb{SS}^T\{\mb{Y}- \bmuhat_0\},\qquad
U_{\beta_G}=\mb{G}^T\{\mb{Y}- \bmuhat_0\},
\\
U_{\tau_I}&=&\{\mb{Y}- \bmuhat_0\}^T
\mathbb{CC}^T\{\mb{Y}- \bmuhat_0\},
\end{eqnarray*}
where $\mb{Y}=({Y_1}, {Y_2},\ldots, {Y_n})^T$,
$\mathbb{X}=(\mb{X}_1, \mb{X}_2,\ldots, \mb{X}_n)^T$,
$\mathbb{S}=(\mb{S}_1, \mb{S}_2,\ldots, \mb{S}_n)^T$,
$\mb{G}=({G_1}, {G_2},\ldots, {G_n})^T$ and
$\mathbb{C}=(\mb{C}_1, \mb{C}_2,\ldots, \mb{C}_n)^T=(G_1\mb{S}_1, G_2\mb
{S}_2,\ldots,\break G_n\mb{S}_n)^T$;
$\bmuhat_0=(\hat\mu_{01},\ldots, \hat\mu_{0n})^T$ and $\hat\mu_{0i}=\exp
(\bX_i^T\hat{\balpha}_0)/\{1+\exp(\bX_i^T\hat{\balpha}_0)\}$ is
the mean of $Y_i$ under $H_0$,
and $\hat{\bs{\alpha}}_0$ is the maximum likelihood estimator of
$\bs{\alpha}$ under the null model
\begin{equation}
\operatorname{logit}\bigl\{P(Y_i=1|\mb{S}_i,
G_i, \mb{X}_i)\bigr\} =\mb{X}_i^T
\bs {\alpha}. \label{nullmodel}
\end{equation}

To combine the three scores to test for the null hypothesis $H_0\dvtx \tau
_S=\tau_I=0$ and $\beta_G=0$, one may consider the conventional score
statistic $Q_{\mathrm{conv}}=\bU^T{\mathcal I}^{-1}\bU$, where $\bU=(U_{\tau
_S},U_{\beta_G}, U_{\tau_I})^T$ and ${\mathcal I}$ is the efficient
information matrix of $\bU$. However, this approach has several major
limitations. First, notice that the score of the regression coefficient
of gene expression $U_{\beta_G}$ is a linear function of $Y$, while the
scores of the variance components of main effects of SNPs and
SNP-by-expression interactions $\tau_S$ and $\tau_I$ are quadratic
functions of $Y$. Hence, $U_{\beta_G}$ has a different scale from
$(U_{\tau_S}, U_{\tau_I})$. It follows that $(U_{\tau_S}, U_{\tau_I})$
are likely to dominate $U_{\beta_G}$. A~combination of the three scores
using $Q_{\mathrm{conv}}$ is hence not desirable. Second, $Q_{\mathrm{conv}}$ involves
quartic functions of $\mb{Y}$ and the information matrix ${\mathcal I}$
involves the 8th moment of $\mb{Y}$. Hence, calculations of $Q_{\mathrm{conv}}$
are not stable, and it is difficult to analytically approximate the
null distribution of $Q_{\mathrm{conv}}$.

We hence propose the following weighted sum of three scores as the test
statistic for the null hypothesis (\ref{null}):
\begin{eqnarray}\label{q-stat}
Q&=&n^{-1}\bigl(a_1 U_{\tau_S}+a_2
U_{\beta_G}^2+a_3 U_{\tau_I}\bigr)
\nonumber\\[-8pt]\\[-8pt]
&=&n^{-1}(\mb{Y}-\bmuhat_0)^T
\bigl(a_1\mathbb{SS}^T+a_2
\mb{GG}^T+a_3\mathbb{CC}^T\bigr) (\mb{Y}-
\bmuhat_0),\nonumber
\end{eqnarray}
where $a_1, a_2, a_3$ are some weights. $Q$ is a nice quadratic
function of $\bY$. Hence, its null distribution can be easily
approximated by a mixture of $\chi^2$ distributions. Different weights
can be chosen. With an equal weight $a_1=a_2=a_3$, $Q$ is equivalent to
the variance component test for $\tau_{\mathrm{common}}$ by assuming $\beta
_{Sj}$, $\beta_G$ and $\gamma_j$ follow a common distribution with mean
zero and variance $\tau_{\mathrm{common}}$. However, this common distribution
assumption is strong, as $\bS$, $G$ and $\bC$ generally have different
scales and so do their effects $\beta_{Sj}$, $\beta_G$ and $\gamma_j$.

Notice\vspace*{-1pt} that $U_{\tau_S}, U_{\beta_G}^2, U_{\tau_I}$ are all quadratic
functions of $\bY$ in similar forms; we propose to weight\vspace*{1pt} each term
$U_{\tau_S}, U_{\beta_G}^2, U_{\tau_I}$ using the inverse of the square
root of their corresponding variances. This allows each weighted term
to have variance 1 and be comparable. Specifically, the variances for
$U_{\tau_S}, U_{\beta_G}^2, U_{\tau_I}$ are
${I}_{\tau_S}=\mb{1}^T(\mathbb{SS}^T \cdot\mathbb{K}\cdot\mathbb
{SS}^T) \mb{1}$, ${I}_{G}=\mb{1}^T(\mb{GG}^T \cdot\mathbb{K}\cdot\mb
{GG}^T) \mb{1}$, ${I}_{\tau_I}=\mb{1}^T(\mathbb{CC}^T \cdot\mathbb
{K}\cdot\mathbb{CC}^T) \mb{1}$, respectively, where $\mathbb{A} \cdot
\mathbb{B}$ denotes the component-wise multiplication of conformable
matrices $\mathbb{A}$ and $\mathbb{B}$, $\mb{1}$ denotes a vector of
ones, the diagonal and off-diagonal elements of $\mathbb{K}$ are
$k_{ii}=-4\hat\mu_{0i}^4+8\hat\mu_{0i}^3-5\mu_{0i}^2+\hat\mu_{0i}$ and
$k_{ii'}=2[\hat\mu_{0i}(1-\hat\mu_{0i})][\hat\mu_{0i'}(1-\hat\mu
_{0i'})]$, respectively [\citet{lin}].

We derive the asymptotic distribution of $Q$ under the null hypothesis
(\ref{null}) by accounting $Q=Q(\hat{\balpha})$ is
a function of $\hat{\balpha}$, which is the maximum likelihood
estimate of $\balpha$ under the null model (\ref{nullmodel}).
Define
\[
\mb{D}= \lleft[\matrix{ \mb{D}_{XX} & \mb{D}_{XV}
\vspace*{3pt}\cr
\mb{D}_{VX} & \mb{D}_{VV} } \rright] =n^{-1}
\mb{U}^T\mb{W}\mb{U},
\]
where $\mb{U}=(\mb{U}_1, \mb{U}_2,\ldots, \mb{U}_n)^T$, $\mb{U}_i= (\mb
{X}_i^T, \mb{V}_i^T)$, $\mb{V}_i^T=(\sqrt{a_1}\mb{S}_i^T, \sqrt {a_2}G_i, \sqrt{a_3}\mb{C}_i^T)$, $\mb{W}=\operatorname{diag}\{\mu
_i(1-\mu_i)\}$.
We show in Section~3 of the supplementary material [\citet{huang}] that under the null hypothesis (\ref{null}), the score test
statistic $Q$ converges in distribution to $Q(0)=\sum_{l=1}^{2p+1}(\mb
{A}_{l}^T\bs{\varepsilon})^2$, where $\bs{\varepsilon}$ is a random vector
following $N(\mb{0}$, $\mb{D}$) and $\mb{A}_{l}$ is the $l$th row of
$\mb{A}=[-\mb{D}_{XV}^T\mb{D}_{XX}^{-1}, \mb{I}_{(2p+1) \times
(2p+1)}]$. This means under the null hypothesis $Q$ follows a~mixture
of $\chi^2$ distributions, which can be approximated using a scaled
$\chi^2$ distribution by matching the first two moments [\citet
{satterthwaite}] as $Q\stackrel{.}{\sim}\kappa\chi_{\nu}^2$, where
$\kappa=\operatorname{Var}(Q)/[2\mathrm{E}(Q)]$ and $\nu=2[\mathrm
{E}(Q)]^2/\operatorname{Var}(Q)$, and the expressions of $\mathrm
{E}(Q)$ and $\operatorname{Var}(Q)$ are given in Section~3 of the
supplementary material [\citet{huang}]. Alternatively, one can
approximate the mixture of $\chi^2$ distribution using the
characteristic function inversion method [\citet{davies}].

\subsection{The omnibus test for the total effect of a gene}\label{sec3.2}
\label{somnibus}

So far we derive the test statistic $Q$ under the outcome model
specified in (\ref{ymodel}), which assumes the disease risk of $Y$
depends on SNPs, gene expression and their interactions. Denote this
$Q$ by $Q_{\mathit{SGC}}$. Suppose that the disease risk of $Y$ depends on SNPs
and gene expression but not their interactions, or even depends only on
SNPs, then it is more powerful to test for the total SNP effect using
the\vspace*{1pt} test statistics $Q_{SG}=n^{-1}(\mb{Y}-\bmuhat_0)^T(a_1\mathbb
{SS}^T+a_2\mb{GG}^T)(\mb{Y}-\bmuhat_0)$ and $Q_S=n^{-1}(\mb{Y}-\bmuhat
_0)^T(a_1\mathbb{SS}^T)(\mb{Y}-\bmuhat_0)$, respectively. Under these
simpler disease models, the test statistic $Q_{\mathit{SGC}}$ loses power as it
tests for extra unnecessary parameters. On the other hand, if the
disease risk indeed depends on SNPs, expression and their interactions,
performing tests only using SNPs $Q_S$ or main effects $Q_{SG}$ will
lose power, compared to $Q_{\mathit{SGC}}$.

Since in reality we do not know the underlying true disease model, it
is difficult to choose a correct model. It is hence desirable to
develop a test that can accommodate different disease models to
maximize the power. Moreover, in a genome-wide association study, it is
almost impossible that one disease model is true for tens of thousands
of genes. Thus, we further propose an omnibus test where we identify
the strongest evidence among the three different models with: (1) only
SNPs, (2)~SNPs and gene expression, and (3) SNPs, gene expression and
their interactions. Specifically, we calculate the $p$-value under each
of the three models, then compute the minimum of the three $p$-values and
compare the observed minimum $p$-value to its null distribution. Because
of the complicated correlation among $Q_{\mathit{SGC}}$, $Q_{SG}$ and $Q_S$, it
is difficult to analytically derive the null distribution of the
minimum $p$-value. To this end, we resort to a resampling perturbation procedure.

As shown in Section~\ref{ssvct}, $Q$ converges in distribution to
$Q(0)=\sum_{l=1}^{2p+1}(\mb{A}_{l}^T\bs{\varepsilon})^2$. The empirical
distribution of $Q(0)$ can be estimated using the resampling method via
perturbation [\citet{parzen}; \citet{cai}]. The
perturbation method approximates the target distribution of $Q$ by
generating random variables $\hat{\bs{\varepsilon}}$ from the
estimated asymptotic distribution of $\bs{\varepsilon}$. This perturbation
procedure is also called the score-based wild bootstrap [\citet{kline}].

Specifically, set $\hat{\bs{\varepsilon}}=n^{-1/2}\sum_{i=1}^n\mb
{U}_i^T(Y_i-\hat\mu_i)\mathcal{N}_i$, where $\mathcal{N}_i$'s are
independent $N(0,1)$. By generating independent $\mathcal{N}=(\mathcal
{N}_1,\ldots,\mathcal{N}_n)$ repeatedly, the distribution of $\hat{\bs{\varepsilon}}$ conditional on the observed data is asymptotically the
same as that of~$\bs{\varepsilon}$. Denoted by $\{\widehat{Q}(0)^{(b)},
b=1,\ldots,B\}$, where $B$ is the number of perturbations, it follows that
the empirical distribution of the $ \widehat{Q}(0)^{(b)}$ is the same
as that of $Q(0)$ asymptotically. The $p$-value can be approximated using
the tail probability by comparing $\{\widehat{Q}(0)^{(b)}, b=1,\ldots,B\}$
with the observed $Q$. Hence, one can calculate the $p$-values of
$Q_{\mathit{SGC}}$, $Q_{SG}$ and $Q_{S}$ by setting $\mb{V}_i=(\sqrt{a_1}\mb
{S}_i^T, \sqrt{a_2}G_i, \sqrt{a_3}\mb{C}_i^T)^T$, $(\sqrt{a_1}\mb
{S}_i^T, \sqrt{a_2}G_i)$ and $\sqrt{a_1}\mb{S}_i^T$ and generate their
perturbed realizations of the null counterpart as $\{\widehat
{Q}_{\mathit{SGC}}(0)^{(b)}\}$, $\{\widehat{Q}_{SG}(0)^{(b)}\}$ and $\{\widehat
{Q}_{S}(0)^{(b)}\}$, and compare them with corresponding observed
values, respectively. Note that for each perturbation $b$, the random
normal perturbation variable $\mathcal{N}^{(b)}$ is the same across the
three tests such that the correlation among $Q_{\mathit{SGC}}$, $Q_{SG}$ and
$Q_S$ can be preserved.

The $p$-value of the omnibus test can be easily calculated using the
perturbation method. Let $\widehat{P}_S=\mathcal{S}_{S}(Q_S)$, $\widehat
{P}_{SG}=\mathcal{S}_{SG}(Q_{SG})$ and\vspace*{1pt} $\widehat{P}_{\mathit{SGC}}=\mathcal
{S}_{\mathit{SGC}}(Q_{\mathit{SGC}})$ be the three $p$-values using the three statistics,
where $\mathcal{S}_{S}(q)=\operatorname{pr}\{\widehat
{Q}_{S}(0)^{(b)}>q\}$, $\mathcal{S}_{SG}(q)=\operatorname{pr}\{\widehat
{Q}_{SG}(0)^{(b)}>q\}$ and $\mathcal{S}_{\mathit{SGC}}(q)=\operatorname{pr}\{
\widehat{Q}_{\mathit{SGC}}(0)^{(b)}>q\}$. The null distribution of the minimum
$p$-value, $\widehat{P}_{\min}=\min(\widehat{P}_S, \widehat{P}_{SG},
\widehat{P}_{\mathit{SGC}})$, can be approximated by $\widehat{P}_{\min
}^{(b)}=\min\{\mathcal{S}_{S}(\widehat{Q}_{S}(0)^{(b)})$, $\mathcal
{S}_{SG}(\widehat{Q}_{SG}(0)^{(b)}), \mathcal{S}_{\mathit{SGC}}(\widehat
{Q}_{\mathit{SGC}}(0)^{(b)})\}$ ($b=1,\ldots, B)$. The $p$-value of the omnibus
test hence can be calculated by comparing the observed minimum $p$-value
$\widehat{P}_{\min}$ with its empirical null distribution $\{\widehat
{P}_{\min}^{(b)}\}$.

Note that different from permutation where the observed data are
shuffled and resampled to calculate the test statistic $Q$, the
perturbation procedure resamples from the asymptotic null distribution
of $Q$ without recalculating $Q$ using the shuffled data. Thus, it is
much more efficient than the permutation method. Using a~\mbox{single} CPU
(2.53 GHz) to run 100 genes (each with 10 SNPs) and 100~cases/100
controls, the computation time is 134.10 and 809.76 seconds for the
perturbation and permutation methods (both with 200 resampling),
respectively. Furthermore, covariates can be easily adjusted using the
perturbation method, but \mbox{covariate} adjustment is more difficult using
permutation. Specifically, the permutation-based $p$-values calculated by
simply permuting SNPs and gene expression fail if SNPs/gene expression
are correlated with covariates.

\section{Understanding the total effect of a gene and the assumptions
of the test using the causal mediation analysis framework}\label{sec4}
\label{sinterpret}
\subsection{Characterization of SNPs, gene expression and disease risk
in the framework of causal mediation modeling}\label{sec4.1}
\label{scmm}

To understand the null hypothesis of no total effect of a gene captured
by SNPs in a gene and a gene expression and the underlying assumptions,
we discuss in this section how to interpret the null in~(\ref{null})
within the causal mediation analysis framework. Causal interpretation
can be helpful for understanding genetic etiology of diseases as well
as for applications in pharmaceutical research [\citet{li2010}].
Although genotype is essentially fixed at \mbox{conception}, it is at that
time effectively randomized, conditional on parental genotypes, and
could be viewed as subject to have hypothetical intervention. The
statistical problem of jointly modeling a set of SNPs, a gene
expression and a~disease can be presented as a causal diagram
[\citet{pearl}; \citet{robins2003}] in \mbox{Figure}~\ref{fig1} and be framed
using a causal mediation model [\citeauthor{vanderweele2009} (\citeyear{vanderweele2009,vanderweele2010});
\citet{imai}] based on counterfactuals [\citeauthor{rubin1974} (\citeyear{rubin1974,rubin1978})]. \citet{vanderweele2010}
and \citet{imai} have used the causal mediation analysis
for epidemiological and social science studies, respectively, where the
exposure of interest is univariate.

One can decompose the \textit{total effect} (TE) of SNPs into the \textit
{Direct Effect} (DE) and the \textit{Indirect Effect} (IE). The \textit{Direct Effect} of SNPs is the effect of the SNPs on the disease outcome
that is not through gene expression, whereas the \textit{Indirect Effect}
of the SNPs is the effect of the SNPs on the disease outcome that is
through the gene expression. Within the causal mediation analysis
framework, we derive in the supplementary material the TE, DE and IE of
the SNPs on the disease outcome [\citet{huang}].

Specifically, we define the total effect (TE) of SNPs as
\[
\mathrm{TE}=\operatorname{logit}\bigl\{P(Y=1|\bS=\mb{s}_1, \bX)\bigr\}-
\operatorname {logit} \bigl\{P(Y=1|\bS=\mb{s}_0, \bX)\bigr\},
\]
that is, the equation (\ref{ymodel}) marginalizes over gene expression
$G$. In Section~1 of the supplementary material [\citet{huang}],
we show that for rare diseases, the total effect of the SNPs on the log
odds ratio (OR) of disease risk
can be expressed in terms of the regression coefficients in models
(\ref{ymodel}) and (\ref{gmodel}) and is approximately equal to
\begin{eqnarray}
\mathrm{TE}&=&(\mb{s}_1-\mb{s}_0)^T\bigl\{\bs{
\beta_S}+\beta_G\bs{\delta}+\bs{\gamma }\bigl(
\mb{x}^T\bs{\phi}+ \mb{s}_0^T\bs{\delta}
\label{te}+ \beta_G\sigma^2_G\bigr)+
\bs{\delta}\mb{s}_1^T\bs{\gamma}\bigr\}
\nonumber\\[-8pt]\\[-8pt]
&&{}+ \tfrac{1}{2}\sigma_G^2(\mb{s}_1+
\mb{s}_0)^T\bs{\gamma}(\mb{s}_1-\mb
{s}_0)^T\bs{\gamma}.\nonumber
\end{eqnarray}

We can express the DE and IE of the SNPs on the log odds ratio of
disease risk in terms of the regression coefficients in models (\ref
{ymodel}) and (\ref{gmodel}). For rare diseases, they are, respectively,
approximately equal to
\begin{eqnarray}
\mathrm{DE}&=&(\mb{s}_1-\mb{s}_0)^T\bigl[
\bs{\beta}_S+\bs{\gamma}\bigl(\mb{x}^T\bs {\phi}+
\mb{s}_0^T\bs{\delta}+ \beta_G
\sigma^2_G\bigr)\bigr]
\nonumber\\[-8pt]\label{de}\\[-8pt]
&&{}+\tfrac{1}{2}\sigma_G^2(\mb{s}_1+
\mb{s}_0)^T\bs{\gamma}(\mb{s}_1-
\mb{s}_0)^T\bs{\gamma},\nonumber
\\
\mathrm{IE}&=&(\mb{s}_1-\mb{s}_0)^T\bs{
\delta}\bigl(\beta_G+\mb{s}_1^T\bs {\gamma}
\bigr). \label{ie}
\end{eqnarray}
These are derived in Section~1 of the supplementary materials using
counterfactuals under the assumptions of no unmeasured confounding
[\citet{huang}].

The sum of the direct and indirect effects is equal to the total effect
of the SNPs, that is, $\mathrm{TE}=\mathrm{DE}+\mathrm{IE}$. As shown
in the supplementary material [\citet{huang}] and discussed in
Section~3.2, identification of the total effect requires a much weaker
assumption than those required for the direct and indirect effects.

\subsection{Understanding the null hypothesis for eQTL~SNPs}\label{sec4.2}
\label{ssnull}

Under the assumption that the gene expression $G$ is associated with
the SNPs $\bS$ (i.e., eQTL SNPs), that is, $\bs{\delta}\neq0$,
using equations (\ref{de}) and (\ref{ie}), the test for the joint
effects of SNPs in a SNP set $\bS$ and a gene expression $G$ on $Y$,
that is, the total effect of a gene, is equivalent to a test for the
total SNP effect on the outcome ($Y$). In fact, for eQTL SNPs, which
have nonzero effects on expression $G$ (i.e., $\bs{\delta}\neq0$),
using the expressions of DE and IE in (\ref{de}) and (\ref{ie}), one
can show that the null hypothesis of no direct and indirect genetic
(SNP) effects is equivalent to the null hypothesis~(\ref{null}) that
all the regression coefficients ($\bs{\beta}_S$, $\beta_G$ and $\bs
{\gamma}$) equal zero:
\[
H_0\dvtx  \bs{\beta}_S=\mb{0},\qquad \beta_G=0,\qquad
\bs{\gamma}=\mb{0}\quad\Leftrightarrow\quad H_0\dvtx \mathrm{DE}=0,\qquad \mathrm{IE}=0.
\label{null0}
\]
%

The null hypothesis (\ref{null}) that all the regression coefficients
($\bs{\beta}_S$, $\beta_G$ and $\bs{\gamma}$) are equal to zero is also
equivalent to the null hypothesis of no total effect of the SNPs
provided $\bs{\beta_S}+\beta_G\bs{\delta}\neq0$ if $\bs{\beta}_S$ or
$\beta_G$ is not 0 for eQTL SNPs ($\delta\neq0)$, that is,
\begin{equation}
\qquad H_0\dvtx  \bs{\beta}_S=\mb{0},\qquad \beta_G=0,\qquad
\bs{\gamma}=\mb{0}\quad\Leftrightarrow\quad H_0\dvtx \mathrm{TE}=\mathrm{DE}+
\mathrm{IE}=0. \label{null1}
\end{equation}

We show in Section~1.4 of the supplementary material that the null
hypothesis~(\ref{null1}) requires only the assumption of no unmeasured
confounding for the effect of eQTL SNPs ($\bS$) on the outcome ($Y$)
after adjusting for the covariates ($\bX$) [\citet{huang}]. Most
genetic association studies make this assumption. In other words, we
make no stronger assumption than standard SNP only analyses for testing
the null hypothesis of no total effect of the SNP set in a~gene.

Note that in models (\ref{ymodel}) and (\ref{gmodel}) we allow other
covariates ($\mb{X}$) to affect both the gene expression and the
disease. If the covariates $\mb{X}$ affect both expression and disease,
ignoring $\mb{X}$ may cause confounding in estimating DE and IE. As
shown in Figure~\ref{fig1}, if arrows from $\mb{X}$ to $G$ and $Y$ exist and $\mb
{X}$ is not controlled for, assumption~(2) in Section~1.2 of the
supplementary material is violated [\citet{huang}]. But if the
covariates $\mb{X}$, the common causes of expression and disease, do
not affect the SNPs $\mb{S}$ (no arrow from $\mb{X}$ to $\mb{S}$), the
estimation and hypothesis testing for TE is still valid. However, if
there does exist an effect of $\mb{X}$ on $\mb{S}$, then it violates
the above assumption of no unmeasured confounding for the $\bS$--$Y$
association and, thus, the test or estimation for TE will be biased.

\subsection{Understanding the null hypothesis for non-eQTL SNPs}\label{sec4.3}

If SNPs have no effect on gene expression ($\bs{\delta}=\mb{0}$), that
is, they are not eQTL SNPs, then there is no indirect effect of the
SNPs on $Y$, so that the null hypothesis of no total effect of a gene
($H_0\dvtx  \bs{\beta}_S=\mb{0}, \beta_G=0, \bs{\gamma}=\mb{0}$) is not
equivalent to testing for no total SNP effect on $Y$ ($H_0\dvtx \mathrm
{TE}=\mathrm{DE}+\mathrm{IE}=0$). In this case, what the null
hypothesis, $H_0\dvtx  \bs{\beta}_S=\mb{0}, \beta_G=0, \bs{\gamma}=\mb{0}$,
tries to evaluate is simply whether there exists a joint effect of the
given set of SNPs $\bS$ and the given gene expression $G$, and possibly
their interactive effect, on disease risk. To test for such a joint
effect, we need the first two assumptions regarding no unmeasured
confounding in Section~1.2 of the supplementary material: no unmeasured
confounding of the SNPs on the outcome and no unmeasured confounding of
the gene expression on the outcome [\citet{huang}].

\subsection{Understanding the traditional genetic analysis using the SNP only model}\label{sec4.4}
\label{ssgwas}

In standard genetic association analysis, we usually fit the following
SNP only model:
\begin{equation}
\operatorname{logit}\bigl\{P(Y_i=1|\mb{S}_i,
\mb{X}_i)\bigr\}=\mb{X}_i^T\bs{\alpha }^*+
\mb{S}_i^T\bs{\beta}_S^*, \label{gwas}
\end{equation}
which does not take gene expression into account, but simply considers
the association between the outcome and SNPs adjusting for covariates.
Note for the special case where SNP, $S$, is univariate, the model (\ref
{gwas}) corresponds to single SNP analysis, the most common approach in
GWAS. \citet{kwee} and \citet{wu} have developed tests for a
SNP-set for $H_0\dvtx \bbeta_S^*=0$ under (\ref{gwas}), which can be more
powerful than individual SNP tests for the association between the
joint effects of the SNPs in a gene and the outcome by borrowing
information across SNPs within a gene, especially when the SNPs are in
good linkage disequilibrium (LD).

Assuming the true models that depend on both SNPs and a gene expression
are specified in (\ref{ymodel}) and (\ref{gmodel}), we study in this
section how $\bs{\beta}_S^*$ in the misspecified standard SNP only
model (\ref{gwas}) is related to the regression parameters $\bs{\beta
}_S, \beta_G$ and~$\bs{\gamma}$ in the true model (\ref{ymodel}) and
what the null hypothesis $H_0\dvtx \bs{\beta}_S^*=0$ under (\ref{gwas})
tests for. To focus on the fundamental issues and for simplicity, we
first discuss the case of no interaction effect between SNPs and gene
expression on disease risk, that is, $\bs{\gamma}=\mb{0}$ in model
(\ref{ymodel}). Under the true $[Y|\bS,G,\bX]$ and $[G|\bS,\bX]$
models in (\ref{ymodel}) and (\ref{gmodel}) assuming no $\bS\times G$
interaction ($\bs{\gamma}=\mb{0}$), by plugging (\ref{gmodel}) into~(\ref{ymodel}), the true $[Y|\bS,G,\bX]$ model can be rewritten as
$\operatorname{logit}\{P(Y_i=1|\mb{S}_i, \mb{X}_i, \varepsilon_i)\}=\mb
{X}_i^T(\bs{\alpha}+\beta_G\bs{\phi})+
\mb{S}_i^T(\bs{\beta}_S+\beta_G\bs{\delta})+\beta_G\varepsilon_i$.
Integrating out $\varepsilon_i\sim N(0,\sigma_G^2)$, we have the true
$[Y|\bS,\bX]$ model as
\begin{equation}
\operatorname{logit}\bigl[P(Y_i=1|\mb{S}_i,
\mb{X}_i)\bigr]\approx c \bigl\{\mb {X}_i^T(
\bs{\alpha}+ \beta_G\bs{\phi})+\mb{S}_i^T(
\bs{\beta}_S+\beta_G\bs{\delta}) \bigr\}, \label{trueysmodel}
\end{equation}
where $c=(1+0.35\times\sigma_G^2\beta_G^2)^{-1/2}$ [\citet{zeger}].

A comparison of (\ref{gwas}) with (\ref{trueysmodel}) shows that $\bs
{\beta}_S^*\approx c(\bs{\beta}_S+\beta_G\bs{\delta})$ and that the
effect of $\mb{S}=\mb{s}_1$ versus $\mb{s}_0$ on the outcome $Y$ under
the SNP only model (\ref{gwas}) corresponds to $(\mb{s}_1-\mb{s}_0)^T\{
c(\bs{\beta}_S+\beta_G\bs{\delta})\}$, which is proportional to the
Total Effect of SNPs in (\ref{te}) when $\bs{\gamma}=\mb{0}$. It
follows that testing for $\bs{\beta}_S^*=\mb{0}$ in the SNP only model~(\ref{gwas}) is approximately equivalent to testing for no total effect
of the SNPs.

However, if there exists a SNP-by-expression interaction on $Y$ and the
SNPs are eQTL SNPs, the naive SNP only analysis using (\ref{gwas}) does
not provide obvious correspondence to the total SNP effect. As shown in
Section~2 of the supplementary material [\citet{huang}], the
induced true $[Y|\bS,\bX]$ model in this setting follows
\begin{eqnarray}\label{trueysint}
&& \operatorname{logit}\bigl\{P(Y_i=1|\mb{S}_i,
\mb{X}_i)\bigr\}
\nonumber\\[-8pt]\\[-8pt]
&&\qquad \approx c_{i}^* \bigl\{\mb{X}_i^T(
\bs{\alpha}+\bs{\phi }\beta_G)+\mb{S}_i^T(
\bs{\beta}_S+\bs{\delta}\beta_G) +\mb{X}_i^T
\bs {\phi}\mb{S}_i^T\bs{\gamma}+\mb{S}_i^T
\bs{\delta}\mb{S}_i^T\bs{\gamma } \bigr\},
\nonumber
\end{eqnarray}
where $c_{i}^*=\{1+0.35\sigma_G^2(\beta_G+\mb{S}_i^T\bs{\gamma})^2\}^{-1/2}$.
This implies that if the $[Y|\mb{S}, G, \mb{X}]$ follows the
interaction model (\ref{ymodel}), the induced true $[Y|\mb{S}, \mb{X}]$
model depends not only on the linear terms of $\mb{X}$ and $\mb{S}$ but
also on the cross-product terms of $\mb{X}$ and $\mb{S}$ and the second
order term of $\mb{S}$. A comparison of (\ref{gwas}) with (\ref
{trueysint}) shows that the standard SNP only model (\ref{gwas})
misspecifies the functional form of the true $[Y|\mb{S}, \mb{X}]$. The
test for $H_0\dvtx  \bs{\beta}_S^*=\mb{0}$ under the misspecified SNP only
model (\ref{gwas}) will still be valid for testing the total effects of
SNPs, because under the null the two models are the same. However, the
misspecified model is subject to power loss, compared to the test based
on the correctly specified model. With only an interaction effect ($\bs
{\gamma}\neq\mb{0}$, $\bs{\beta}_S=\mb{0}$, $\beta_G=0$), (\ref
{trueysint}) can be\vspace*{1pt} written as $\operatorname{logit}[P(Y_i=1|\mb{S}_i,
\mb{X}_i)]
\approx c_{i}^* \{\mb{X}_i^T\bs{\alpha}
+\mb{X}_i^T\bs{\phi}\mb{S}_i^T\bs{\gamma}+\mb{S}_i^T\bs{\delta}\mb
{S}_i^T\bs{\gamma} \},
$ where $c_{i}^*=\{1+0.35\sigma_G^2(\mb{S}_i^T\bs{\gamma})^2\}^{-1/2}$.
If we assume this is the true model and fit the conventional GWAS model
(\ref{gwas}) to test for the SNP effect, the test is again still valid
under the null, but loses power under the alternative.

\subsection{Understanding the relation with Mendelian randomization}\label{sec4.5}

The approach here differs in several ways from that based on Mendelian
randomization [Smith and Ebrahim (\citeyear{smith2003} and \citeyear
{smith2005})] in which genetic markers (SNPs) are instrumental
variables to assess the effect of an exposure (in our case, a gene
expression value) on an outcome. Here we are interested in using a gene
expression to increase power for testing for the total effect of SNPs
on a disease outcome. Furthermore, Mendelian randomization makes the
assumption that SNPs do not have an effect on an outcome except through
an exposure (e.g., gene expression in our case), in\vadjust{\goodbreak} other words, no
direct effect. No such assumption is being made here. This is because
we are interested in testing for a different effect, that is, the
effect of SNPs, rather than the effect of an exposure (gene expression)
on disease risk.

\section{Simulation studies}\label{sec5}
\label{ssimulation}

\subsection{Simulation setup}\label{sec5.1}

To make the simulation mimic the motivating asthma data [\citet
{moffatt}], we simulated data using the \textit{ORMDL3} gene on
chromosome 17q21. We generated the SNP data in the \textit{ORMDL3} gene
by accounting for its linkage disequilibrium structure using HAPGEN
based on the CEU sample [\citet{marchini}]. The genomic location
used to generate the SNP data is between 35.22 and 35.39 Mb on
chromosome 17, which contains 99 HapMap SNPs. Ten of the 99 HapMap SNPs
are genotyped on the Illumina HumanHap300 array, that is, 10 typed SNPs.

To generate gene expression and the disease outcome, we assumed there
is one causal SNP $S_{\mathrm{causal}}$ and varied the causal SNP among the 99
HapMap SNPs in each simulation. In Section~\ref{sec5.4} we further perform a
simulation study assuming three causal SNPs. For subject $i$, gene
expression $G_i$ was generated by the linear regression model
$G_i=0+\delta\times S_{\mathrm{causal}, i}+\varepsilon_{i}$, $\varepsilon_{i}\sim N(0,
1.44)$. The outcome $Y_i$ was generated by the logistic model
\begin{eqnarray*}
&& \operatorname{logit}\bigl\{P(Y_i=1|S_{\mathrm{causal}, i},
G_i)\bigr\}
\\
&&\qquad =-0.2+\beta_S\times S_{\mathrm{causal}, i}+
\beta_G\times G_i+\gamma\times G_iS_{\mathrm{causal}, i}.
\end{eqnarray*}
The parameters and the range of $\beta_S$, $\beta_G$ and $\gamma$ were
based on the empirical estimates from analysis of the asthma data. For
each simulation, we first generated a cohort with 1000 subjects, and
100 cases and 100 controls were randomly selected from the 1000
subjects to form a case--control sample.

Two sets of simulations were performed. In the first set, we selected
the SNP rs8067378 as the causal SNP, as this SNP is highly associated
with asthma in the original GWAS [\citet{moffatt}]. For each
configuration of $\beta_S$, $\beta_G$, $\gamma$ and $\delta$, we
generated 2000 data sets to calculate the empirical size and power. In
the second set of simulation, the causal SNP was chosen one at a time
out of the 99 HapMap SNPs. For each selected causal SNP, we generated
1000 data sets and evaluated statistical power for two different
disease models: $(\beta_S, \beta_G, \gamma)=(0.4, 0, 0)$ or $(0.2, 0.2,
0)$. In both simulation settings, we used the 10 typed SNPs of the
\mbox{\textit{ORMDL3}} gene on the Illumina chip to form the SNP-set for the
model (\ref{ymodel}), that is, $p=10$, in calculating the test
statistics $Q_S$, $Q_{SG}$, $Q_{\mathit{SGC}}$ and the omnibus test. For
$Q_{SG}$ and $Q_{\mathit{SGC}}$, both weighted and unweighted methods were
investigated, where $a_1=1$, $a_2=(I_{G}/I_{\tau_S})^{-1/2}$ and
$a_3=(I_{\tau_I}/I_{\tau_S})^{-1/2}$ for the weighted statistic, and
$a_1=a_2=a_3=1$ for the unweighted statistic. The $p$-values were
calculated using the scaled $\chi^2$ approximation, the Davies' method
by inverting the characteristic function [\citet{davies}] and the
perturbation procedure with 500 perturbations. The results of these
approximations were very similar at the\vadjust{\goodbreak} significance level of 0.05. We
performed the omnibus test by combining the evidence from $Q_S$,
weighted $Q_{SG}$ and weighted $Q_{\mathit{SGC}}$.

\subsection{Size and power: By varying effect sizes for a fixed causal SNP}\label{sec5.2}
\label{ssinglesnp}
We first evaluated the sizes of the proposed score tests, where the
null distribution was approximated by either the scaled $\chi^2$
approximation or the perturbation procedure (Table~\ref{tab1}). Type I errors
are well protected using both approximation methods under the three
models with statistics $Q_S, Q_{SG}, Q_{\mathit{SGC}}$. The empirical size is
close to 0.05 for the omnibus test and the three models.
As the results using different approximation methods are similar at the
level of 0.05, we only present in the following the empirical power
using the perturbation method. We also evaluate the performance of the
proposed tests using the characteristic function inversion method
[\citet{davies}] and the perturbation method at smaller sizes
($\alpha=5\times10^{-3}$ and $5\times10^{-4}$) (Table~\ref{tab1} of the
supplementary material [\citet{huang}]), and find the methods
perform well.

%
\begin{table}[b]
\tabcolsep=0pt
\caption{Empirical sizes (\%) of the proposed tests using scaled $\chi^2$
approximation and the perturbation. The size was calculated at the
significance level of 0.05 based on 2000 simulations}\label{tab1}
\begin{tabular*}{\tablewidth}{@{\extracolsep{\fill}}@{}lcccc@{}}
\hline
& \multicolumn{2}{c}{\textbf{Scaled $\bolds{\chi^2}$ approximation}}
& \multicolumn{2}{c}{\textbf{Perturbation}}
\\[-6pt]
& \multicolumn{2}{c}{\hrulefill} & \multicolumn{2}{c}{\hrulefill}
\\
& \textbf{Unweighted} & \textbf{Weighted} & \textbf{Unweighted} & \textbf{Weighted}
\\
\hline
\multicolumn{5}{c}{Gene expression depends on SNPs}\\
SNPs & \multicolumn{2}{c}{4.65} & \multicolumn{2}{c}{4.90}\\
SNPs and expression & 4.80 & 4.95 & 4.80 & 4.80 \\
SNPs, expression and interaction & 4.75 & 4.60 & 4.60 & 4.35\\
Omnibus test & \multicolumn{3}{c}{--} & 5.15
\\[3pt]
\multicolumn{5}{c}{Gene expression and SNPs are independent}\\
SNPs & \multicolumn{2}{c}{4.83} & \multicolumn{2}{c}{4.97}\\
SNPs and expression & 4.87 & 4.90 & 4.83 & 4.87 \\
SNPs, expression and interaction & 4.60 & 4.80 & 4.77 & 5.07 \\
Omnibus test & \multicolumn{3}{c}{--} & 5.13 \\
\hline
\end{tabular*}
\end{table}

%
\begin{figure}
\begin{tabular}{@{}cc@{}}
\includegraphics{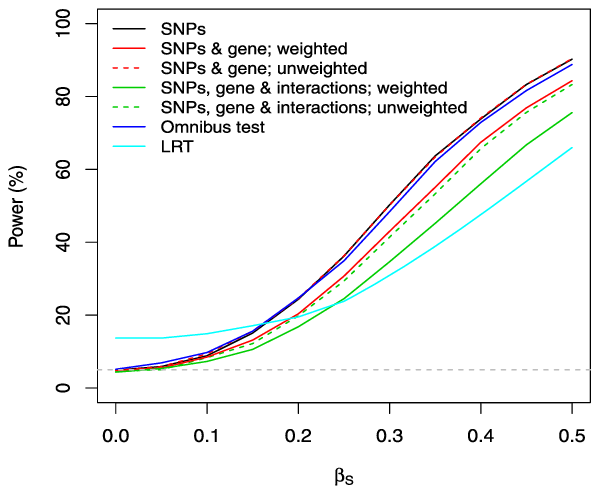} & \includegraphics{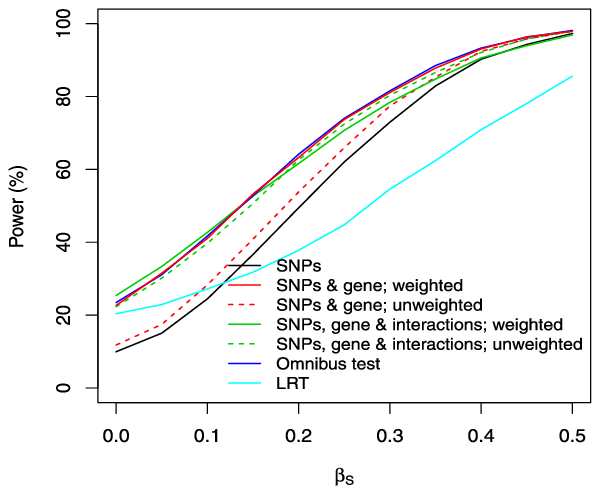}\\
\scriptsize{(a) $\beta_G=0, \gamma=0; \delta=1.0$} & \scriptsize{(b) $\beta_G=0.2, \gamma=0; \delta=1.0$}
\end{tabular}\vspace*{6pt}
\includegraphics{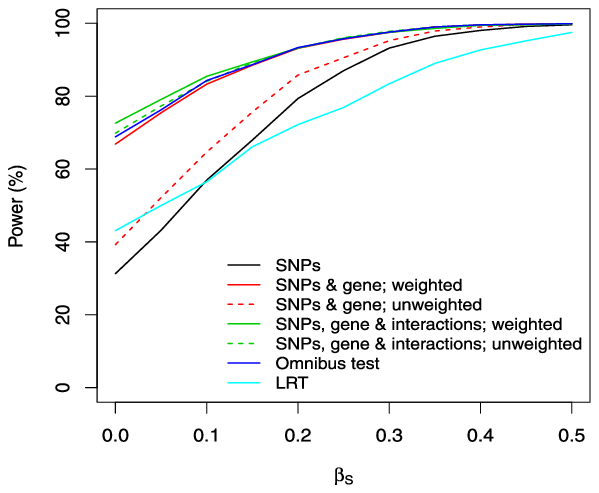}
\scriptsize{(c) $\beta_G=0.2, \gamma=0.2; \delta=1.0$}
\caption{Empirical power. SNPs are assumed to be eQTL SNPs $(\delta=1)$. Each
figure plots the powers of the proposed tests as a function of the main
effect of the SNP ($\beta_s)$. The three figures correspond to the
three different true models, the model with only SNP effects, the model
with main effects without interaction and the model with SNPs, gene
expression and their interaction effects. The dashed line in \textup{(a)}
indicates 5\% type I error rate.}\label{fig2}
\end{figure}

We assumed rs8067378 is the causal SNP and eQTL, and compared the
powers of the three statistics $Q_S$, $Q_{SG}$ and $Q_{\mathit{SGC}}$ as well as
the omnibus test under three different configurations of effect sizes
(Figure~\ref{fig2}). The first setting assumes both gene expression and the
interactions between the SNPs and the gene expression have no effect on
the outcome ($\beta_G=0$ and $\gamma=0$) [Figure~\ref{fig2}(a)]. As expected,
$Q_S$ shows the optimal power as it correctly specifies the model. The
other two tests $Q_{SG}$ and $Q_{\mathit{SGC}}$, especially $Q_{\mathit{SGC}}$,
over-specify the model and waste DF for testing the effects of
expression and interactions, and hence lose power. The performance of
the omnibus test is close to the $Q_S$. As a note, the correctly
specified model here means that gene expression or nonlinear
interaction has been incorporated in the analyses, not that typed SNPs
are causal.

The second setting assumes the gene expression has an effect on the
outcome but there is no interaction ($\beta_G=0.2$ and $\gamma=0$)
[Figure~\ref{fig2}(b)], while the third setting further assumes an interaction
effect ($\beta_G=0.2$ and $\gamma=0.2$) [Figure~\ref{fig2}(c)]. The tests
$Q_{SG}$ and $Q_{\mathit{SGC}}$, respectively, have the best power under the
correct model. In contrast to the setting 1, $Q_S$ has the worst
performance among the three tests when expression has an effect on
disease, and the power loss of $Q_S$ is even more in the presence of
the interactions [Figure~\ref{fig2}(c)]. The unweighted $Q_{SG}$ also does not
perform well in these two cases and has considerable loss of power. The
rest of the tests have similar power. The performance of the omnibus
test in both settings is close to the optimal test with only minimal
loss of power.

We also study in Figure~\ref{fig2} the performance of the likelihood ratio test
(LRT) for testing for the joint effects of SNPs and gene expression by
comparing the model with an intercept, SNPs, gene expression and
interactions with the model with only the intercept. In general, our
proposed methods outperform the LRT in both power and type I error. The
power loss and the incorrect size of the LRT are likely due to the
large degrees of freedom relative to the sample size (DF${}={}$21; 100
cases/100 controls) and the high LD among some of the typed SNPs.

\subsection{Power: By varying causal SNPs}\label{sec5.3}
\label{smultiplesnp}
\begin{figure}[t]
\includegraphics{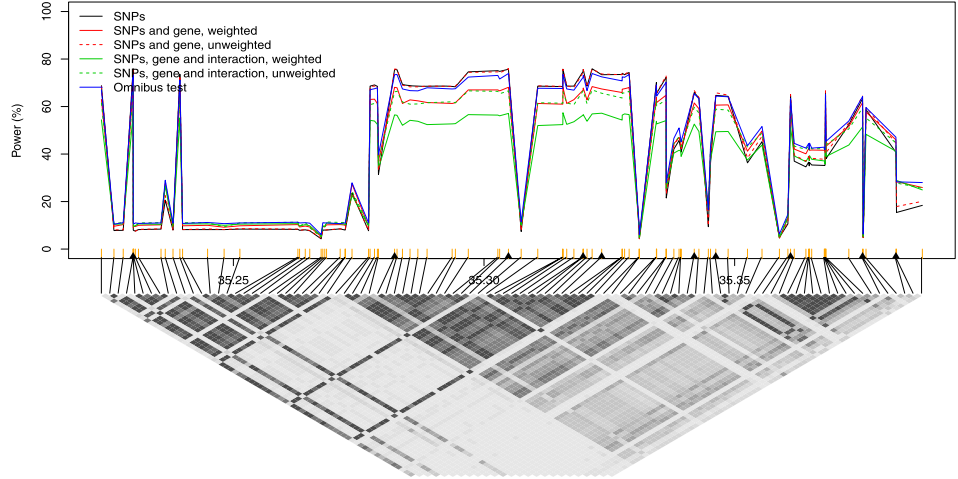}
\scriptsize{(a) $\beta_S=0.4, \beta_G=0, \gamma=0$}\vspace*{6pt}
\includegraphics{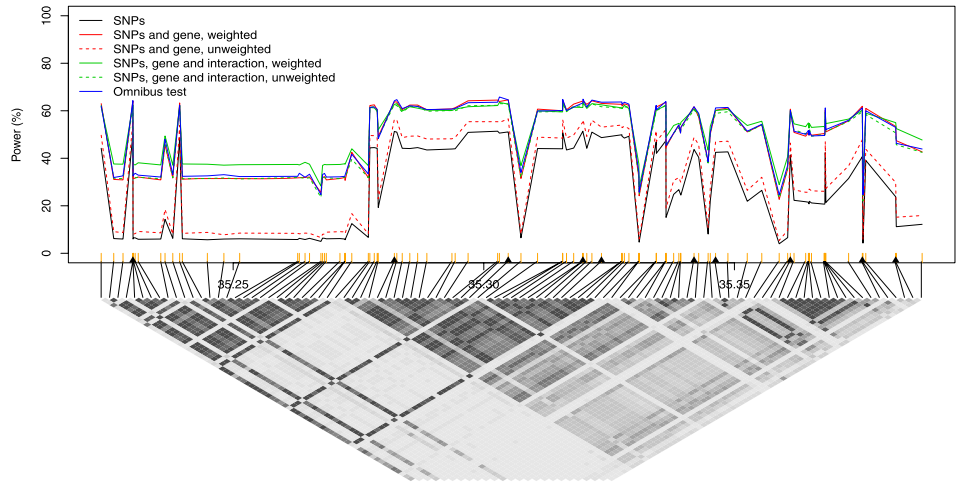}
\scriptsize{(b) $\beta_S=0.2, \beta_G=0.2, \gamma=0$}
\caption{Simulated power curves for evaluating how different choices of causal
SNPs affect the powers of the proposed tests. The x-axis indicates the
physical location (Mb) of the 99 HapMap SNPs at 17q21. The orange
vertical bar indicates the relative locations of the causal SNP and the
black triangles indicate the ten typed SNPs. Different lines indicate
the power\vspace*{1pt} of different tests. The lower panel of each subfigure is the
plot for linkage disequilibrium, measured as $r^2$ ranging from 0~(white) to 1 (black).}\label{fig3}
\end{figure}

In order to investigate the performance under ``synthetic
association,'' that is, the causal variant is untyped (i.e., not on a
chip) [\citet{dickson}], we assessed the power of the methods when
each of the 99 HapMap SNPs was assumed to be causal. In particular, we
were interested in evaluating how the correlation between the causal
SNP and the 10 typed SNPs affected the statistical powers of the
proposed tests. Intuitively, if the causal SNP is untyped and has low
LD with the typed SNPs, one would expect lower power. We considered two
settings: the outcome is only associated with the causal SNP: $\beta
_S=0.4$, $\beta_G=0$, $\gamma=0$ [Figure~\ref{fig3}(a)]; and the outcome is
associated with the causal SNP and the gene expression but not their
interaction: $\beta_S=0.2$, $\beta_G=0.2$, $\gamma=0$ [Figure~\ref{fig3}(b)]. A
total of 1000 simulations were performed.

%

Figures~\ref{fig3}(a) and (b) show that the pattern of simulation results is
very similar to those in the previous section. The test assuming the
correct model performs the best. The omnibus test nearly reaches the
optimal power obtained under the true model in both settings. In
addition, the test using the weighted statistic derived under the model
with SNPs and gene expression as predictors (weighted~$Q_{SG}$)
performs well even when the interaction model is true (data not shown),
although it has some loss of power in setting 1 when only SNPs are
associated with the outcome. As the model (\ref{ymodel}) can be written
as $\operatorname{logit}[P(Y_i=1|\mb{S}_i, G_i, \mb{X}_i)]=\mb{X}_i^T\bs
{\alpha}+\mb{S}_i^T\bs{\beta}^*+G_i\beta_G$ where $\bs{\beta}^*=\bs
{\beta}_S+G_i\bs{\gamma}$, the main effect only analyses can still
capture the interactive effect $\gamma$ even though the model is not
correctly specified. So the simpler test $Q_{SG}$ can be used as an
alternative to the omnibus test if the computation cost is a concern.

Figure~\ref{fig3} also shows that statistical power depends on the correlation
between the causal SNP (which might be untyped) and the 10 typed SNPs.
The power rises as the correlation between the causal SNP and the typed
SNPs increases. For example, the statistical power is high if a causal
SNP is in the LD block spanned between the second to the third typed
SNPs (marked as the second and third black triangles from left to
right, according to the physical location, in Figure~\ref{fig3}), as it has good
correlation with the typed SNPs. The power is generally low if a causal
variant lies in the region between the first and second typed SNPs, as
it has little correlation with the typed SNPs. In this case, it is
virtually not possible to detect genetic effects using the typed SNPs
on the chip no matter what method one uses. Although the typed SNPs
might not include the underlying causal SNPs, it still provides a valid
testing procedure due to the same model under the null. However, the
typed SNPs may or may not provide a consistent estimate for the effect
of the causal SNP, depending on the LD pattern of the causal SNP and
the typed~SNPs.

\subsection{Additional simulations: Model misspecification, multiple
causal variants, varying LD structures}\label{sec5.4}
\label{smore}

\begin{figure}[b]
\begin{tabular}{@{}cc@{}}
\includegraphics{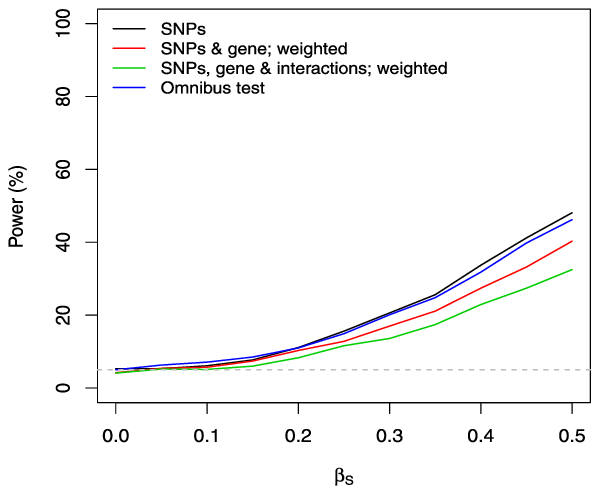}& \includegraphics{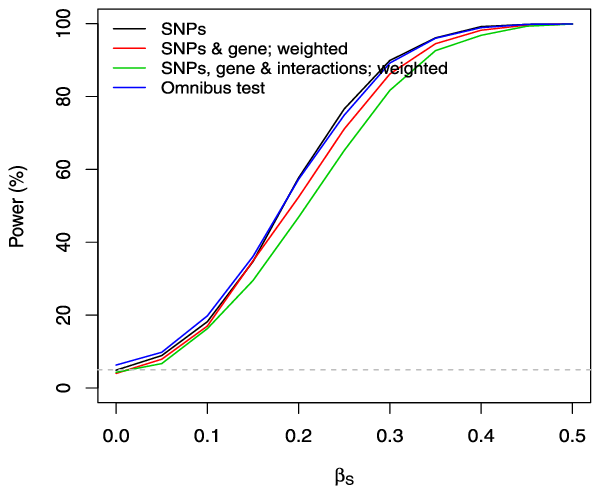}\\
\scriptsize{(a) $\beta_G=0, \gamma=0; \delta=1.0$} & \scriptsize{(b) $\beta_G=0, \gamma=0; \delta=1.0$}\\
\includegraphics{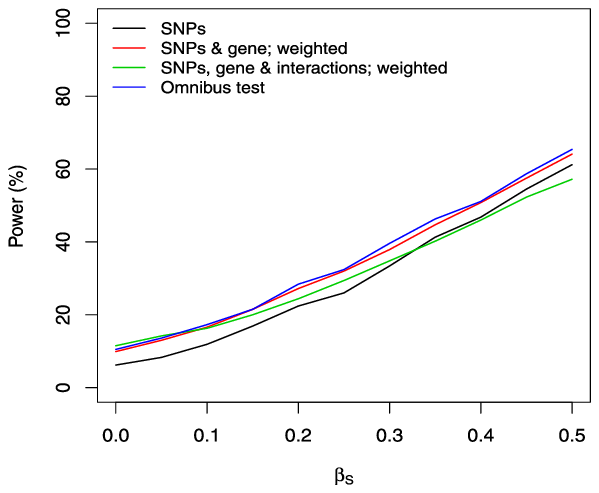}
&
\includegraphics{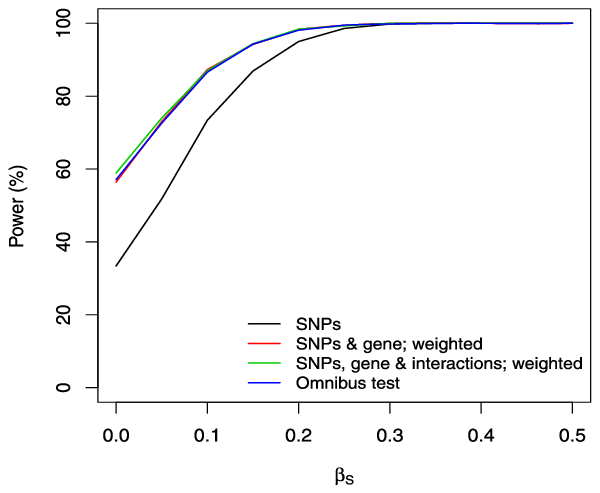}\\
\scriptsize{(c) $\beta_G=0.2, \gamma=0; \delta=1.0$} & \scriptsize{(d) $\beta_G=0.2, \gamma=0; \delta=1.0$}
\end{tabular}
\caption{Empirical power under model misspecification. SNPs are assumed
to be eQTL SNPs $(\delta=1)$. Each figure plots the powers of the
proposed tests as a function of the main effect of SNP ($\beta_s)$. The
six figures correspond to the different true models: the model with
only SNP effects [\textup{(a)}~and~\textup{(b)}], the model with main effects of SNP and
gene expression [\textup{(c)} and \textup{(d)}] and the model with SNPs, gene expression
and their interaction effects [\textup{(e)} and \textup{(f)}]. \textup{(a)}, \textup{(c)}, \textup{(e)} are
simulated under $\operatorname{logit}[P(Y_i=1|S_{\mathrm{causal}, i},
G_i)]=-100^{0.9}+(100+\beta_S S_{\mathrm{causal}, i}+\beta_G G_i+\gamma
G_iS_{\mathrm{causal}, i})^{0.9}$ and \textup{(b)}, \textup{(d)}, \textup{(f)} are simulated under the
probit model $\Phi^{-1}[P(Y_i=1|S_{\mathrm{causal}, i}, G_i)]=-0.2+\beta_S
S_{\mathrm{causal}, i}+ \beta_G G_i+\gamma G_iS_{\mathrm{causal}, i}$. The dashed lines in
\textup{(a)} and \textup{(b)} indicate 5\% type I error rate.}\label{fig4}
\end{figure}
%
\begin{figure}
\begin{tabular}{@{}cc@{}}
\includegraphics{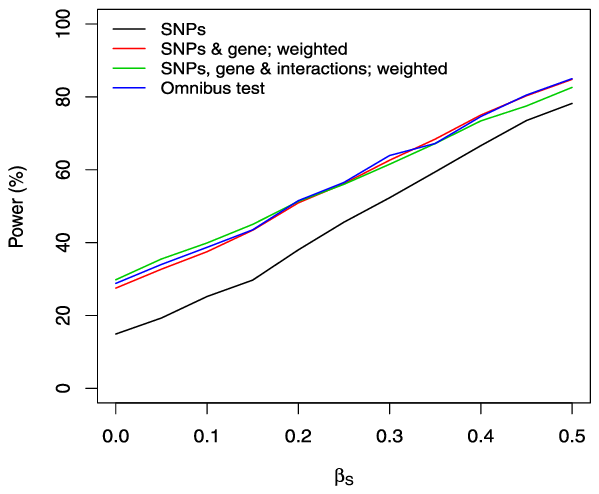} & \includegraphics{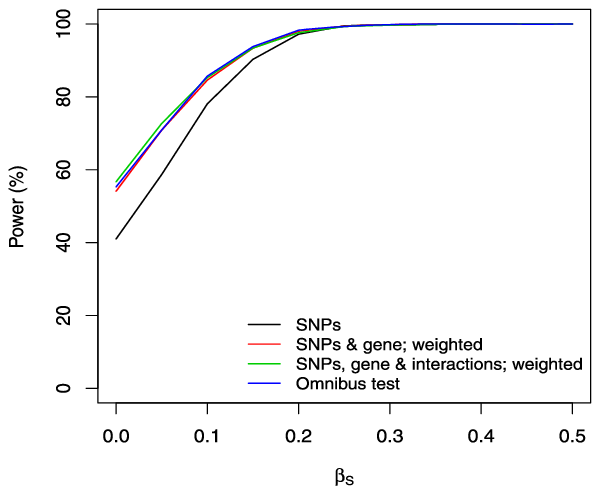}\\
\scriptsize{(e) $\beta_G=0.2, \gamma=0.2; \delta=1.0$} & \scriptsize{(f) $\beta_G=0.05, \gamma=0.15; \delta=1.0$}
\end{tabular}
\caption{(Continued).}
\end{figure}

We performed additional simulations to assess how model
misspecification influences our proposed test. Gene expression $G_i$
was generated without the normality assumption $G_i=0+\delta S_{\mathrm{causal},
i}+\varepsilon_{i}$, $\varepsilon_{i}\sim N(0, 1.44)+\operatorname{uniform}(-0.3, 0.3)$. Two
outcome models are explored. The first model generates the outcome
$Y_i$ by the logistic model assuming nonlinear effects of SNPs and $G$ as
$\operatorname{logit}\{P(Y_i=1|S_{\mathrm{causal}, i}, G_i)\}
=-100^{0.9}+(100+\beta_S S_{\mathrm{causal}, i}+\beta_G G_i+\gamma G_iS_{\mathrm{causal},
i})^{0.9}$, and the second model generates $Y_i$ by a probit model $\Phi
^{-1}\{P(Y_i=1|S_{\mathrm{causal}, i}, G_i)\}=-0.2+\beta_S S_{\mathrm{causal}, i}+\beta_G
G_i+\gamma G_iS_{\mathrm{causal}, i}$. Although the model is not correctly
specified in our proposed test under these settings, the joint analyses
of SNPs and expression still outperform the SNP-only analyses when the
gene expression contributes to the risk of developing disease.
Similarly, the performance of the omnibus test is very close to the
optimal test obtained under the true model for different scenarios
(Figure~\ref{fig4}).

We conducted two additional simulation studies by varying the number of
causal variants and LD structures. The pattern of the results from
these additional studies is very similar to what is presented above
(Figures~\ref{fig1} and~\ref{fig2} in the supplementary material [\citet{huang}]).
The first additional study is similar to the study in Section~\ref{ssinglesnp}, except that there are three causal SNPs in the \textit{ORMDL3} gene instead of a single causal SNP. Using the same ten typed
SNPs for the analyses, we again found that the test performs the best
when the model is correctly specified and the omnibus test approaches
the optimal test obtained under the true model with limited power loss
(Figure~\ref{fig1} of the supplementary material [\citet{huang}]).

Similar to analyses in Section~\ref{smultiplesnp}, the second
additional study investigates the performance of the proposed test at
15q24--15q25.1, where SNPs have a different LD pattern from the \textit{ORMDL3} gene. Assuming one causal SNP at a time, we used the same ten
typed SNPs to perform our proposed test. Again, the test performs the
best when the model is correctly specified, and the omnibus test is
robust and approaches the optimal test obtained by assuming the true
model, and the power depends on the correlation of the causal SNP and
the typed SNPs (Figure~\ref{fig2} of the supplementary material [\citet{huang}]).

\section{Analysis of the asthma data}\label{sec6}
\label{sdataanalysis}

We applied the proposed testing procedures to reinvestigate the genetic
effects of
the \textit{ORMDL3} gene on the risk of childhood asthma in the MRC-A
data [\citet{dixon}; \citet{moffatt}]. This subset of the
data contained 108 asthma cases and 50 controls where we have complete
data of the 10 typed SNPs and gene expression of \textit{ORMDL3}. The
SNP data were genotyped using the Illumina 300K chip and the gene
expression was collected using the Affymetrix Hu133A 2.0. We analyzed
the data using both additive and dominant modes: in the additive mode,
the genotype was coded as the number of the minor allele (i.e., 0, 1,
2), whereas in the dominant mode, the genotype was coded as whether or
not the minor allele was present (i.e., 0, 1).

%
\begin{table}[b]
\tabcolsep=0pt
\caption{$p$-values of the effects of the 10 typed SNPs at \textit{ORMDL3}
on the risk of childhood asthma. Different rows indicate the predictors
to be tested. $P_{\min}$ calculates the minimum $p$-value using
individual SNP analyses; VCT is the proposed variance component test}\label{tab2}
\begin{tabular*}{\tablewidth}{@{\extracolsep{\fill}}@{}lccccc@{}}
\hline
& \textbf{Multivariate} & \textbf{Bonferroni-}                 & \textbf{Permutation-}                & \textbf{VCT}        &  \textbf{VCT}\\
& \textbf{Wald}         & \textbf{adjusted $\bolds{P_{\min}}$} & \textbf{adjusted $\bolds{P_{\min}}$} & \textbf{unweighted} &  \textbf{weighted}
\\
\hline
\multicolumn{6}{c}{Additive model}\\
SNPs & 0.122 & 0.102 & 0.039 & \multicolumn{2}{c}{0.044}\\
SNPs, gene & 0.342 & 0.194 & 0.057 & 0.039\phantom{0} & 0.033\phantom{0}\\
SNPs, gene and interaction & 0.013 & 0.303 & 0.093 & 0.0025 & 0.0028 \\
Omnibus test & \multicolumn{3}{c}{--} & \multicolumn{2}{c}{0.0055}
\\[3pt]
\multicolumn{6}{c}{Dominant model}\\
SNPs & 0.018 & 0.015 & 0.018 & \multicolumn{2}{c}{0.0045}\\
SNPs, gene & 0.094 & 0.031 & 0.018 & 0.0040 & 0.0040 \\
SNPs, gene and interaction & 0.131 & 0.098 & 0.048 & 0.0035 & 0.0023 \\
Omnibus test & \multicolumn{3}{c}{--} & \multicolumn{2}{c}{0.0030}\\
\hline
\end{tabular*}
\end{table}

We applied the proposed tests for the total SNP effect of \textit
{ORMDL3} using the SNP and gene expression data. There are strong
associations between the SNPs and the gene expression (8 out of 10 with
$p$-value${}<{}$0.05 and the other two with $p$-values of 0.076 and 0.21), that
is, the SNPs are eQTL SNPs. We considered six test statistics: $Q_S$,
unweighted $Q_{SG}$, weighted $Q_{SG}$, unweighted $Q_{\mathit{SGC}}$, weighted
$Q_{\mathit{SGC}}$, and omnibus test (Table~\ref{tab2}). We compared our methods with the
standard multivariate or univariate methods: the multivariate Wald
test, which has 10, 11 and 21 degrees of freedom under the three models
(SNPs only, SNPs and gene expression, and SNPs, gene expression and
interactions). We also included in the comparison the test using the
smallest $p$-value from the 10 single SNP analyses with the Bonferroni
adjustment or the adjustment using the permutation procedure to account
for the correlation among the SNPs.

The results in Table~\ref{tab2} show that our proposed methods give smaller
$p$-values compared to the standard testing procedures. The test
$Q_{\mathit{SGC}}$, which accounts for the effects of SNPs, gene expression and
their interactions, gives the smallest \mbox{$p$-}value compared to the tests
only using SNPs, in both additive and dominant modes. For example, the
$p$-values using weighted $Q_{\mathit{SGC}}$ and $Q_S$ are 0.0028 and 0.044,
respectively, using the additive SNP model. The omnibus test calculated
using the perturbation procedure by computing the minimum $p$-value from
$Q_S$, weighted $Q_{SG}$ and weighted $Q_{\mathit{SGC}}$ also provides a more
significant signal than those only considering SNPs, with the $p$-value
being 0.0055. These results are consistent with the findings in
simulation studies.

We also performed genome-wide analyses for both SNP-sets and single
SNP. We first paired eQTL SNPs with their corresponding gene expression
[\citet{dixon}] and performed SNP-only analyses and our proposed
method. For single SNP analyses, after adjustment of multiple
comparison using false discovery rate [FDR; \citet{storey2002}],
56 SNPs with FDR${}<{}$0.1 were identified in SNP-only analyses and 97
SNPs were identified from the proposed omnibus test. For SNP-set
analyses, we grouped eQTL SNPs that correspond to the same gene as a
SNP-set, and we identified 5 and 15 SNP-sets (FDR${}<{}$0.1) from SNP-only
analyses and omnibus tests, respectively.

\section{Discussion}\label{sec7}
\label{sdiscussion}

We proposed in this paper to integrate SNP and gene expression data to
improve power for genetic association studies. The major contributions
of this paper are as follows: (1) to formulate the data integration
problem of different types of genomic data as a mediation problem; (2)
to propose a powerful and robust testing procedure for the total effect
of a gene contributed by SNPs and a gene expression; and (3) to relax
the assumptions required for mediation analyses in the test for the
total effect.

Specifically, as shown in Figure~\ref{fig1}, we are able to integrate the
information of SNPs and gene expression as a biological process through
the mediation model.
Our proposed variance component score test for the total effect of SNPs
and a gene expression circumvents the instability of estimation of the
joint effects of multiple SNPs and gene expression, because only the
null model needs to be fit. Mediation analysis to estimate direct and
indirect effects generally requires additional unmeasured confounding
assumptions, and previous work mainly focused on estimation. Here we
focus on testing for the total effect of a gene using SNPs and a gene
expression. For eQTL SNPs, we show that the total effect of a gene
contributed by SNPs and a gene expression is equivalent to the total
effect of SNPs, which is the sum of direct and indirect effects of SNPs
mediated through gene expression. Testing for the total SNP effect only
requires one assumption: no unmeasured confounding for the effect of
SNPs on the outcome, which is the same assumption as the standard GWAS
and, thus, no stronger assumption is required.

We characterize the relation among SNPs, gene expression and disease
risk in the framework of causal mediation modeling. This framework
allows us to understand the null hypothesis of no total effect of a
gene contributed by SNPs and gene expression, and the underlying
assumptions of the test for both eQTL SNPs and non-eQTL SNPs. We
propose a variance component score test for the total effects of a gene
on disease. This test allows to jointly test for the effects of SNPs,
gene expression and their interactions. We showed that the proposed
test statistic follows a mixture of $\chi^2$ distributions
asymptotically, and proposed to approximate its finite sample
distribution using a scaled $\chi^2$ distribution, a characteristic
function inversion method or a perturbation method.

We considered three tests: using only SNPs ($Q_S$), SNPs and gene
expression main effects ($Q_{SG}$), and SNPs, gene expression and their
interactions ($Q_{\mathit{SGC}}$). Our simulation study shows that all three
tests have the correct type I error for testing for the overall SNP
effect. The relative power of these tests depends on the underlying
true relation between the predictors (SNPs, gene expression and their
interactions) and disease. As the underlying biology is often unknown,
we further constructed the omnibus test that identifies the most
powerful test among the three disease models, and proposed to use the
perturbation method to calculate the \mbox{$p$-}value for the omnibus test.
Further, the test using only the main effects of SNPs and gene
expression loses limited power compared to the omnibus test and can be
used as a simple alternative.

Our results also show that to test for the total effects of a gene, the
tests that incorporate both SNP and gene expression information, such
as $Q_{SG}$ and $Q_{\mathit{SGC}}$, are more powerful if SNPs are associated
with gene expression than if they are not. In other words, it is even
more beneficial to incorporate gene expression data with SNP data to
detect genetic effects on disease if gene expression is a good causal
mediator for the SNPs. To achieve this, a natural way is to select SNPs
located within or at the neighborhood of a gene, since it has been well
destablished that the SNP within a gene can alter its expression value
via transcription regulation [\citet{lee}]. Alternatively, one
can restrict the joint SNP-expression analysis to known eQTL SNPs. If
selection of eQTL SNPs is based on statistical significance, one also
needs to be aware of the possibility that cis-action may be a
confounding effect of SNPs on array hybridization [\citet{li2006}].

We mainly focus on testing for the total effect of a gene in this
paper. The proposed method can be easily extended to test for direct
and indirect effects separately for eQTL SNPs. Using equation (\ref
{de}), to test for the direct effect of the SNPs, one can test $H_0\dvtx \bs
{\beta}_S=0,\bs{\gamma}=0$. Using the notation in equation (\ref
{q-stat}), one can test this null hypothesis using the statistic
$Q_{\mathrm{DE}}=n^{-1}(a_1 U_{\tau_S}+a_3 U_{\tau_I} )$, where the
null model is a logistic model with $\bs{X}$ and $\bs{G}$. To test for
the indirect effects of the SNPs, using equation (\ref{ie}), one can
test $H_0\dvtx \beta_G=0,\bs{\gamma}=0$. Using the notation in equation (\ref
{q-stat}), one can test this null hypothesis using the statistic
$Q_{\mathrm{IE}}=n^{-1}(a_2 U_{\beta_G}^2+a_3 U_{\tau_I})$, where the
null model is a logistic model with $\bs{X}$~and~$\bs{S}$. As the
number of SNPs $\bs{S}$ might be large and some SNPs might be highly
correlated (with high LD), standard regression to fit the null model
might not work well. One can fit the null model using ridge regression.
To perform these tests, one will need to make the four unmeasured
confounding assumptions required for estimating direct and indirect
effects of SNPs stated in Section~1.2 of the supplementary material
[\citet{huang}].

Gene expression may not be the only mediator for the relation between
SNPs and disease. Other biomarkers, such as DNA methylation, proteins,
metabolites of the gene product in the blood, immunological or
biochemical markers in the serum, and environmental factors can also
serve as potential mediators, depending on the context or the disease
to be studied. For instance, epigenetic variations have been reported
to exert heritable phenotypic effects [\citet{johannes}].
Furthermore, our proposed test can be applied to address many other
scientific questions as long as there exists a causal relationship as
illustrated in Figure~\ref{fig1}. For example, the SNP-gene-disease relations
can be replaced by the DNA copy number-protein-cancer stage (early vs.
late) in tumor genomics studies to assess if copy number can have any
effect on the clinical stage of cancer. It is advantageous to set up a
biologically meaningful model before applying our proposed test
procedure, which makes the best use of the prior knowledge.


\section*{Acknowledgments} The authors would like to thank the Editor,
the Associate Editor and the referees for their helpful comments that
have improved the paper.

\begin{supplement}
\stitle{Detailed causal and statistical development and supplementary table and figures}
\slink[doi]{10.1214/13-AOAS690SUPP} 
\sdatatype{.pdf}
\sfilename{aoas690\_supp.pdf}
\sdescription{Section~1: detailed development of causal mediation model and derivations referenced in Sections~\ref{scmm} and~\ref{ssnull};
Section~2: derivation of model (\ref{trueysint}) in Sections~\ref{ssgwas};
Section~3: asymptotic distribution of $Q$ referenced in Section~\ref{ssvct}; table and
figures referenced in Sections~\ref{ssinglesnp} and~\ref{smore}.}
\end{supplement}



%

\printaddresses

\end{document}